\documentclass[aps,prl,twocolumn,showpacs,showkeys,preprintnumbers]{revtex4}
\usepackage{eurosym}
\usepackage{amsmath}
\usepackage{dcolumn}
\usepackage{bm}
\usepackage{subfigure}
\usepackage{amsfonts}
\usepackage{amssymb}
\usepackage{makeidx}
\usepackage{epsfig}
\usepackage{graphicx}

\begin{document}

\title{\textbf{Kinetic formulation of Tolman-Ehrenfest effect: non-ideal
fluids in Schwarzschild and Kerr space-times}}
\author{Claudio Cremaschini$^{a}$, Ji\v{r}\'{\i} Kov\'{a}\v{r}$^{a}$, Zden%
\v{e}k Stuchl\'{\i}k$^{a}$ and Massimo Tessarotto$^{b,a}$}
\affiliation{$^{a}$Research Centre for Theoretical Physics and Astrophysics, Institute of
Physics, Silesian University in Opava, Bezru\v{c}ovo n\'{a}m.13, CZ-74601
Opava, Czech Republic}
\affiliation{$^{b}$Department of Mathematics and Geosciences, University of Trieste, Via
Valerio 12, 34127 Trieste, Italy\\
}
\date{\today }

\begin{abstract}
A review of the original thermodynamic formulation of the Tolman-Ehrenfest
effect prescribing the temperature profile of uncharged fluid at thermal
equilibrium forming stationary configurations in curved space-time is
proposed. A statistical description based on relativistic kinetic theory is
implemented. In this context the Tolman-Ehrenfest relation arises in the
Schwarzschild space-time for collisionless uncharged particles at Maxwellian
kinetic equilibrium. However, the result changes considerably when non-ideal
fluids, i.e., non-Maxwellian distributions, are treated, whose statistical
temperature becomes non-isotropic and gives rise to a tensor pressure. This
is\ associated with phase-space anisotropies in the distribution function,
occurring both for diagonal and non-diagonal metric tensors, exemplified by
the Schwarzschild and Kerr metrics respectively. As a consequence, it is
shown that for these systems it is not possible to define a Tolman-Ehrenfest
relation in terms of an isotropic scalar temperature. Qualitative properties
of the novel solution are discussed.
\end{abstract}

\maketitle

In General Relativity the Tolman-Ehrenfest effect refers to the physical law
discovered by R. C. Tolman and P. Ehrenfest in 1930, which determines the
temperature profile of a perfect fluid at thermal equilibrium in curved
space-time. More precisely, in its original formulation \cite{Tolman},
Tolman found that in a spherically symmetric and stationary space-time
characterized by metric tensor $g_{\mu \nu }$ with signature $\left(
-,+,+,+\right) $, the equilibrium scalar temperature of a corresponding
spherically symmetric stationary perfect fluid of uncharged (i.e., neutral)
matter is not uniformly constant. Under these conditions, in fact, due to
space-time curvature the temperature becomes a radial-position dependent
function $T\left( \rho \right) $ that satisfies the constraint relation%
\begin{equation}
T_{\ast }=T\left( \rho \right) \sqrt{-g_{00}\left( \rho \right) },
\label{T-1}
\end{equation}%
where $\rho $\ is the radial coordinate in a spherical coordinate system $%
(\varphi ,\vartheta ,\rho \mathbf{)}$ and $T_{\ast }=const.$ represents the
temperature measured at $\rho =\infty $ in the asymptotically-flat frame for
which $g_{00}\left( \infty \right) =-1$. In a subsequent paper published
jointly by Tolman and Eherenfest in 1930 \cite{Ehren} it was proved that an
analogous result can also be established even in absence of spherical
symmetry assumption. This occurs provided there exists an appropriate
reference frame (i.e., a coordinate system) in which the metric tensor $%
g_{\mu \nu }$ is diagonal and both the space-time metric and the fluid are
stationary with respect to the corresponding coordinate-time $t$. In such a
setting, denoting by $\xi ^{\mu }$ the timelike Killing vector field, the
Tolman-Ehrenfest effect can be cast in tensorial representation yielding the
temperature profile $T\left( \mathbf{r}\right) $ at the generic $3-$position 
$\mathbf{r=(}r_{1},r_{2},r_{3}\mathbf{)}$, expressed in arbitrary
coordinates $\mathbf{(}r_{1},r_{2},r_{3}\mathbf{)}$, as%
\begin{equation}
T_{\ast }=T\left( \mathbf{r}\right) \sqrt{g_{\mu \nu }\left( \mathbf{r}%
\right) \xi ^{\mu }\xi ^{\nu }},  \label{2}
\end{equation}%
where again\ $T_{\ast }=const.$

The Tolman-Ehrenfest relation represents a fundamental result of theoretical
physics and relativistic fluid dynamics. It proves that the curvature of
space-time affects the thermodynamic physical properties of fluid systems,
so that the dynamical principles of General Relativity equally apply also to
the heat distribution and transfer in space-time \cite{Rovelli,Visser0}. For
these reasons, the potential interest and practical implementation of the
Tolman-Ehrenfest relation in astrophysics and observational astronomy are of
great interest. In particular, this concerns the study of the dynamics of
compact objects and related relativistic processes of matter accretion and
distribution occurring in their surrounding \cite{pop4}, for which the
temperature profile can represent an observational target \cite{Visser1}. In
these cases, in fact, it is crucial to relate the fundamental parameters of
the metric solution to the intrinsic physical properties of matter sources,
represented by neutral or weakly-ionized gases \cite{gas2,gas3,gas4},
plasmas and fluids \cite{Maha01,pop0}, possibly subject to electromagnetic
or radiation fields \cite{Maha02,pop1,pop2,pop2bis,pop3}. Additional
theoretical applications of the Tolman-Ehrenfest law have been also proposed
in relation to the concept of Unruh temperature and quantum emission
phenomena described by the Hawking effect at black-hole event horizons \cite%
{Gim}. The Tolman-Ehrenfest mathematical relation finds therefore a fertile
field of application. However, it is instructive to quest about the
effective universal character that might be attributed to the same solution,
in order to establish its domain of validity and possible limitations. In
this regard, two main features of the Tolman-Ehrenfest theory must be
examined: a)\ the restriction to ideal relativistic fluids characterized by
isotropic scalar temperature, which excludes non-ideal effects like
configuration-space anisotropies (e.g., viscosity); b) the requisite of
having a diagonal metric tensor.

Given these premises, the purpose of the present research is to carry out an
analytical review of the original thermodynamic formulation of the
Tolman-Ehrenfest effect, with application to non-ideal fluids. To this aim,
a statistical description for the distribution of neutral matter (i.e., a
collisionless system of neutral point particles) in curved space-time is
implemented, based on relativistic kinetic theory \cite{degroot}. Metric
tensor representations corresponding to spherically-symmetric and stationary
solutions of the Einstein field equations are considered. In such a
framework the Tolman-Ehrenfest relation arises for collisionless matter at
Maxwellian kinetic equilibrium in Schwarzschild space-time. However, the
physical setting changes considerably when extension to non-ideal (i.e.,
non-Maxwellian) fluids is carried out, which are characterized by occurrence
of phase-space anisotropies in the distribution function. For these systems
in fact the statistical temperature and corresponding fluid pressure become
intrinsically non-isotropic, requiring a tensorial representation in terms
of directional temperatures. An explicit realization of solutions of this
kind in Schwarzschild and Kerr space-times is considered, showing that for
these systems a Tolman-Ehrenfest relation in terms of an isotropic scalar
temperature of the type (\ref{2}) is not generally admissible.

In order to set the mathematical framework for the target of the study, we
recall basic knowledge of statistical theory and kinetic equilibrium. Two
kinds of statistical descriptions can be formulated for large $N-$body
systems formed by a number $N\gg 1$ of constituent particles. These include
relativistic microscopic kinetic theories \cite{EPJ2,EPJ5}, relying on
particle dynamics and yielding information about the statistical
distribution of matter in particle configuration and velocity spaces, and
fluid theories, according to which instead the $N-$body system is treated as
a continuum in configuration space described by observable fluid fields. For
the case of relativistic neutral collisionless systems treated here the
appropriate statistical framework is provided by the first category, and
more precisely by the covariant Vlasov kinetic theory, to be coupled to the
Einstein field equations. This allows for both phase-space single-particle
as well as gravitational collective system dynamics to be consistently taken
into account. The fundamental quantity is represented by the probability (or
kinetic)\ distribution function (KDF) $f=f\left( \mathbf{x},s\right) $
defined on $\Gamma _{1}\times I$, being $\Gamma _{1}$\ the single-particle
phase-space for the state $\mathbf{x}$ (see below) and $I\subseteq 
\mathbb{R}
$\ the axis of proper-time $s$. Its dynamical evolution is determined by the
Vlasov equation, which in Lagrangian form reads%
\begin{equation}
\frac{d}{ds}f\left( \mathbf{x}\left( s\right) ,s\right) =0,  \label{vlasov-1}
\end{equation}%
where\ $s$ and $\mathbf{x}\left( s\right) \equiv \left( r^{\mu }\left(
s\right) ,u^{\mu }\left( s\right) \equiv \frac{dr^{\mu }\left( s\right) }{ds}%
\right) $ denote respectively the particle proper-time and the Lagrangian
state, with $r^{\mu },u^{\mu }$\ being the particle $4-$position and $4-$%
velocity\ and, as usual, $\mu =0,1,2,3$\ denotes the tensorial index. The
KDF may correspond, in particular, to a kinetic equilibrium, to be denoted $%
f_{\ast }$.\ This means that, when referred to a suitable
general-relativistic reference frame, the KDF is required to be a smooth,
strictly-positive ordinary function which depends only on a suitable set of
invariants $I_{l}(\mathbf{x})$ (with $0<l\leq 3)$, namely such that $\frac{d%
}{ds}I_{l}(\mathbf{x}(s))=0$, where the real scalar functions $I_{l}(\mathbf{%
x})$\ identify\ independent first integrals of motion, i.e., invariant phase
functions which do not explicitly depend on the proper-time $s$.

The kinetic approach yields intrinsically a more fundamental level of
description than the fluid one, whereby the velocity-integrals of the KDF
define appropriate fluid fields, namely physical observables, while the
velocity moments of the kinetic equation (\ref{vlasov-1}) determine the
corresponding set of continuum fluid equations. The issue is relevant in
collisionless systems, since they can develop phase-space anisotropies, or
due to the peculiar form of the KDF itself, which may depart from an
isotropic local Maxwellian function, implying the occurrence of distinctive
collective phenomena \cite{Carter0,Carter1,Carter2}. The statistical,\ i.e.,
kinetic, treatment of these features is therefore essential in order to
characterize the dynamical and thermodynamical properties of the systems of
interest.

For definiteness, in the following we consider the case of a single-species $%
N-$body system composed of neutral particles having unitary rest mass, so
that hereon $m=1$, belonging to a $4-$dimensional smooth Lorentzian manifold 
$\mathcal{R}^{4}$ having metric tensor $g_{\mu \nu }$. Given a generic
particle with $4-$position $r^{\mu }$ and velocity $4-$vector $u^{\mu
}\equiv \frac{dr^{\mu }}{ds}$, the line element $ds$ defining the
proper-time satisfies the identity%
\begin{equation}
ds^{2}=g_{\mu \nu }\left( r\right) dr^{\mu }dr^{\nu },  \label{ds}
\end{equation}%
with $g(r)\equiv \left\{ g_{\mu \nu }\left( r\right) \right\} \equiv \left\{
g^{_{\mu \nu }}\left( r\right) \right\} $ denoting the metric field tensor
evaluated at the $4-$position $r\equiv \left\{ r^{\mu }\right\} $. This
equation implies the \textit{mass-shell constraint} for the $4-$velocity $%
u^{\mu }$, namely%
\begin{equation}
g_{\mu \nu }\left( r\right) u^{\mu }u^{\nu }=1,  \label{mass-shell1}
\end{equation}%
which must be intended as identically satisfied along the particle geodesic
trajectory, which is determined by the equation of motion $\frac{D}{Ds}%
u^{\mu }\left( s\right) =0$, where $\frac{D}{Ds}$ denotes the covariant
derivative. For the covariant representation of the particle dynamics we
adopt a formalism based on the introduction of a tetrad of orthogonal unit $%
4-$vectors $\left( a^{\mu },b^{\mu },c^{\mu },d^{\mu }\right) $, where $%
a^{\mu }$ and $\left( b^{\mu },c^{\mu },d^{\mu }\right) $ are respectively
time-like and space-like. In terms of them the particle $4-$velocity can
therefore be decomposed as%
\begin{equation}
u^{\mu }\equiv u^{\left( a\right) }a^{\mu }+u^{\left( b\right) }b^{\mu
}+u^{\left( c\right) }c^{\mu }+u^{\left( d\right) }d^{\mu },
\label{4-veeldeco}
\end{equation}%
to be denoted as tetrad representation, where $\left( u^{\left( a\right)
},u^{\left( b\right) },u^{\left( c\right) },u^{\left( d\right) }\right) $
identify the corresponding components to be separately identified as $4-$%
scalars, while the tensorial indexes are carried by the unit $4-$vectors.
Then, for single-particle dynamics the unconstrained $8-$dimensional
phase-space $\Gamma _{1}$ is defined as $\Gamma _{1}=\mathcal{R}^{4}\times 
\mathcal{V}^{4}$, where the $4-$dimensional velocity space $\mathcal{V}^{4}$
is defined as the tangent bundle of the configuration-space manifold $%
\mathcal{R}^{4}$ whose $4-$vectors are subject to the mass-shell constraint (%
\ref{mass-shell1}). On the other hand, the tetrad representation provides a
local mutual relationship among the components of the $4-$velocity (\ref%
{4-veeldeco}). Consistent with (\ref{mass-shell1}), this yields the
representation%
\begin{equation}
u^{\left( a\right) }=\sqrt{u^{\left( b\right) }{}^{2}+u^{\left( c\right)
}{}^{2}+u^{\left( d\right) }{}^{2}-1}.  \label{u-zero}
\end{equation}

In terms of this setting, a generic relativistic equilibrium KDF $f_{\ast }$
is said to be isotropic on the velocity space $\mathcal{V}^{4}$ if it
carries even powers of all the $4-$velocity components $\left( u^{\left(
b\right) },u^{\left( c\right) },u^{\left( d\right) }\right) $ and this
functional dependence is isotropic. A particular case of isotropic
dependence is through the velocity component $u^{\left( a\right) }$. An
example is represented by the relativistic Maxwellian function $f_{M}$. In
contrast, a KDF is said to be non-isotropic on $\mathcal{V}^{4}$ if it
exhibits a non-isotropic dependence on the even powers of the $4-$velocity
components $\left( u^{\left( b\right) },u^{\left( c\right) },u^{\left(
d\right) }\right) $. Non-isotropic distribution functions necessarily differ
from local Maxwellian distributions. From a statistical point of view such
deviations are due to velocity-space, or more generally phase-space kinetic
effects arising from single-particle dynamics, which are characteristic of
collisionless systems. In such a framework, the existence of non-isotropic
KDFs is associated with the occurrence of corresponding non-ideal fluids.

In order to address this feature for the study of the Tolman-Ehrenfest
effect and compare with ideal fluids, it is necessary to investigate the
structure of the stress-energy tensor $T^{\mu \nu }\left( r\right) $
associated with the equilibrium KDF $f_{\ast }$ and defined by the $4-$%
velocity integral%
\begin{equation}
T^{\mu \nu }\left( r\right) =2\int_{\mathcal{V}^{4}}\sqrt{-g}d^{4}u\Theta
\left( u^{0}\right) \delta \left( u^{\mu }u_{\mu }-1\right) u^{\mu }u^{\nu
}f_{\ast },  \label{tmunu}
\end{equation}%
where the Dirac-delta takes into account the kinematic constraint (\ref%
{mass-shell1}) for the $4-$velocity when performing the integration, $\Theta 
$ denotes the Theta-function selecting the root of the time-component $u^{0}$%
, while $\sqrt{-g}$ is the square-root of the determinant of the metric
tensor. Invoking the tetrad representation for the $4-$velocity (\ref%
{4-veeldeco}), the integral (\ref{tmunu}) can be reduced to%
\begin{equation}
T^{\mu \nu }\left( r\right) =\int \frac{\sqrt{-g}d^{3}u}{\sqrt{u^{\left(
b\right) }{}^{2}+u^{\left( c\right) }{}^{2}+u^{\left( d\right) }{}^{2}-1}}%
u^{\mu }u^{\nu }f_{\ast },  \label{tmunu-bis}
\end{equation}%
which is defined over the $3-$dimensional tangent space in which the
component $u^{\left( a\right) }$ of the $4-$velocity depends on the other
components according to Eq.(\ref{u-zero}),\ while for the locally-flat
tangent velocity space $\sqrt{-g}=1$. The result is generally a non-trivial
solution for $T^{\mu \nu }\left( r\right) $, whose explicit evaluation of
the tensorial components requires numerical integration. Nevertheless, for
the purpose of the present research we shall proceed with an analytical
analysis of the qualitative features of $T^{\mu \nu }\left( r\right) $.

Having set the required formalism, we proceed with the study of the
Tolman-Ehrenfest effect. We first consider the case in which the metric
tensor $g(r)$\ corresponds to a space-time generated by a central
non-rotating mass $M$ identified with the\ (axisymmetric) Schwarzschild
solution. Adopting spherical-like coordinates $r\equiv \left( t,\phi ,\rho
,\theta \right) $ and the geometrical system of units ($c=G=1$), the line
element is diagonal and is written as%
\begin{equation}
ds^{2}=-\left( 1-\frac{\rho _{s}}{\rho }\right) dt^{2}+\left( 1-\frac{\rho
_{s}}{\rho }\right) ^{-1}d\rho ^{2}+\rho ^{2}d\Omega ,  \label{Schwar}
\end{equation}%
where $\rho _{s}\equiv 2M$ is the Schwarzschild radius and $d\Omega \equiv
d\theta ^{2}+\sin ^{2}\theta d\phi ^{2}$ is the solid-angle element. In the
same coordinate system we denote the covariant components of the particle $%
4- $velocity as $\left( u_{t},u_{\phi },u_{\rho },u_{\theta }\right) $.
Thanks to the stationarity and axisymmetry assumption of the Schwarzschild
metric tensor $g(r)$, the coordinates $t$ and $\phi $ are by construction
ignorable for the same metric tensor. This implies that there are two
Killing vectors, denoted as $\xi ^{\alpha }=\delta _{t}^{\alpha }$ and $%
\zeta ^{\alpha }=\delta _{\phi }^{\alpha }$, which in turn determine also
the corresponding integrals of motion to be identified respectively with the
total particle energy $E$ and angular momentum $L$. These are given by%
\begin{eqnarray}
E &\equiv &-\xi ^{\mu }u_{\mu }=\left[ 1-\frac{\rho _{s}}{\rho }\right] 
\overset{\cdot }{t},  \label{E-Schwar} \\
L &\equiv &\zeta ^{\mu }u_{\mu }=\rho ^{2}\sin ^{2}\theta \overset{\cdot }{%
\phi },  \label{L-Schwar}
\end{eqnarray}%
where $\overset{\cdot }{t}=\frac{dt}{ds}\equiv u^{t}$ and $\overset{\cdot }{%
\phi }=\frac{d\phi }{ds}\equiv u^{\phi }$.

The realization of a relativistic kinetic equilibrium for the collisionless $%
N-$body system of neutral matter in Schwarzschild space-time\ can then be
obtained by adopting the general relativistic kinetic approach developed in
Refs.\cite{pop0,pop1,pop2}. This is reached by expressing the KDF in terms of%
\textbf{\ }the single-particle invariants, identified here with the set $%
I\left( \mathbf{x}\left( s\right) \right) =(E,L)$ and defined respectively
by Eqs.(\ref{E-Schwar}) and (\ref{L-Schwar}). The simplest choice is to
retain exponential linear functions of the invariants and to represent the
equilibrium KDF in the form%
\begin{equation}
f_{\ast }=f_{\ast }\left( E,L,\Lambda _{\ast }\right) ,  \label{f-star}
\end{equation}%
assumed to be a smooth strictly-positive function which is summable in
velocity-space. Concerning the notation, in Eq.(\ref{f-star})\ the set $%
\left( E,L\right) $ denotes explicit functional dependences\ carried by the
KDF on the invariants, while $\Lambda _{\ast }$ denotes the so-called
structure functions \cite{pop0}, namely possible additional equilibrium
dependences that are related to the observable velocity moments of the KDF
(i.e., the physical fluid fields of the system). A realization includes the
case in which $\Lambda _{\ast }$ are identically constant, namely $\Lambda
_{\ast }=const.$ The precise form of the equilibrium KDF can be constructed
based on the adoption of the principle of maximum entropy \cite{Jaynes}.
Following analogous treatments carried out for relativistic collisionless
plasmas \cite{pop1,pop2}, it is possible to introduce an explicit
representation for the KDF consistent with Eq.(\ref{f-star}) in the form of
a relativistic drifted-Maxwellian distribution letting $f_{\ast }=f_{M}$,
where%
\begin{equation}
f_{M}=\beta _{\ast }e^{-E\gamma _{\ast }-L\omega _{\ast }}.
\label{equil-gaussian}
\end{equation}%
Here, $\Lambda _{\ast }=\left( \beta _{\ast },\gamma _{\ast },\omega _{\ast
}\right) $ are uniform fluid fields, where $\beta _{\ast }$ is associated
with the particle number density measured in the fluid comoving frame, $%
\gamma _{\ast }\equiv \frac{1}{T_{\ast }}$ determines the system isotropic
temperature $T_{\ast }$, while $\omega _{\ast }$ defines the fluid angular
frequency along the direction $\phi $ of spatial symmetry, when measured by
an inertial observer. The Maxwellian KDF is an isotropic distribution and
realizes an ideal fluid configuration. For relativistic collisionless
stationary and spherically-symmetric systems at kinetic equilibrium
described by a Maxwellian KDF of the type (\ref{equil-gaussian}) the
Tolman-Ehrenfest effect applies in the form of Eq.(\ref{T-1}). The proof of
the statement can be found in Ref.\cite{gravi}. Under these circumstances
therefore the thermodynamic temperature and the statistical kinetic
temperature are isotropic and exhibit the same profile affected by the
curvature of space-time.

However, it is important to point out that more complicated functional
dependences of the equilibrium KDF on the particle invariants are in
principle admitted in the Schwarzschild case. This freedom in fact is
allowed, in the framework of the Vlasov theory for collisionless systems, to
obtain consistent equilibrium solutions. As a result, even preserving the
diagonal form of the space-time metric tensor, this yields a class of
non-isotropic solutions for the equilibrium KDFs that differ from the
Maxwellian distribution considered in Eq.(\ref{equil-gaussian}) and for
which the Tolman-Ehrenfest effect (\ref{T-1}) does not generally apply. More
precisely, two distinct effects can be identified in this respect:

A)\ The first one is related to the functional form of the structure
functions, that are in principle allowed to carry implicit dependences on $%
I\left( \mathbf{x}\left( s\right) \right) $, and therefore to be of the
general form%
\begin{equation}
\Lambda _{\ast }=\Lambda _{\ast }\left( E,L\right) .
\end{equation}%
In kinetic theory, the inclusion in the KDF of this functional dependences
on the particle invariants can be treated analytically in terms of a
Chapman-Enskog expansion holding in appropriate kinetic regimes \cite{pop0}.
Thus, if one introduces a suitable dimensionless expansion parameter $%
\varepsilon \ll 1$, to first order the equilibrium KDF can finally be
written as%
\begin{equation}
f_{\ast }=f_{o}\left( 1+\varepsilon f_{1}\right) +O\left( \varepsilon
^{2}\right) ,
\end{equation}%
where $f_{o}$ is the leading-order equilibrium solution, while the
first-order correction $f_{1}$ carries information about the KDF phase-space
gradients associated with the occurrence of non-uniform fluid fields. An
example is represented by energy or temperature gradients of the non-ideal
fluid across equipotential surfaces associated with the Schwarzschild
space-time. In detail, if one sets for simplicity $f_{o}\equiv f_{M}$, then
the leading-order temperature is isotropic. In such a case one can write
similarly%
\begin{equation}
T^{\mu \nu }\left( r\right) =T_{o}^{\mu \nu }\left( 1+\varepsilon T_{1}^{\mu
\nu }\left( r\right) \right) +O\left( \varepsilon ^{2}\right) ,
\end{equation}%
where for the Maxwellian case $T_{o}^{\mu \nu }=T_{o}\mathbf{I}$, with $%
T_{o}=const.$ and $\mathbf{I}$ being the unit diagonal tensor. However, the
integrals (\ref{tmunu}) over $f_{1}$ are sufficient to produce a
physically-meaningful deviation from the isotropic configuration, generating
a diagonal stress-energy tensor $T_{1}^{\mu \nu }\left( r\right) $ for the
non-ideal fluid, with a corresponding temperature profile violating the
Tolman-Ehrenfest law.

B)\ The second effect consists in assuming that the equilibrium KDF depends
on some power $n$ of the invariants, with $n\geq 2$. An interesting test
solution of this type is%
\begin{equation}
f_{\ast }=f_{\ast }\left( E,L,L^{2}\right) ,
\end{equation}%
with explicit realization in terms of the Gaussian KDF%
\begin{equation}
f_{M}=\beta _{\ast }e^{-E\gamma _{\ast }-L\omega _{\ast }-L^{2}\eta _{\ast
}}.
\end{equation}%
Again, the solution can be treated by means of a Chapman-Enskog expansion,
assuming now that the contribution $L^{2}\eta _{\ast }$ is of $O\left(
\varepsilon \right) $ for a suitable kinetic regime. The result is a
non-ideal fluid characterized by a temperature anisotropy carried by the
tangential component $T^{\phi \phi }\left( r\right) $ differing from $%
T^{\rho \rho }\left( r\right) =T^{\theta \theta }\left( r\right) $. Then,
since the corresponding directional temperatures have different spatial
profiles, the Tolman-Ehrenfest relation (\ref{T-1}) cannot be established.

Let us now consider the realization of the Tolman-Ehrenfest effect in the
field of an uncharged rotating black hole of rest mass $M$ and intrinsic
angular momentum $S$ described by the Kerr metric. In Boyer-Lindquist
coordinates $\left( t,\phi ,\rho ,\theta \right) $ and geometrical system of
units the corresponding metric is expressed in terms of the non-diagonal
line element as%
\begin{eqnarray}
ds^{2} &=&-\left[ \frac{\Delta -a^{2}\sin ^{2}\theta }{\Sigma }\right]
dt^{2}-\frac{4Ma\rho \sin ^{2}\theta }{\Sigma }dtd\phi  \notag \\
&&+\left[ \frac{\left( \rho ^{2}+a^{2}\right) ^{2}-\Delta a^{2}\sin
^{2}\theta }{\Sigma }\right] \sin ^{2}\theta d\phi ^{2}  \notag \\
&&+\frac{\Sigma }{\Delta }d\rho ^{2}+\Sigma d\theta ^{2},  \label{ker}
\end{eqnarray}%
where $\Sigma \equiv \rho ^{2}+a^{2}\cos ^{2}\theta $ and $\Delta \equiv
\rho ^{2}+a^{2}-2M\rho $, with $a\equiv S/M$ being the angular momentum per
unit mass. Like the Shwarzschild solution, also the Kerr metric admits two
Killing vectors associated with the ignorable coordinates $t$ and $\phi $.
This implies that the total particle energy $E$ and angular momentum $L$ are
integrals of motion, given by%
\begin{equation}
E\equiv -\xi ^{\mu }u_{\mu }=\left[ 1-\frac{2M\rho }{\Sigma }\right] \overset%
{\cdot }{t}+\frac{2Ma\rho \sin ^{2}\theta }{\Sigma }\overset{\cdot }{\phi },
\label{ker-e}
\end{equation}%
\begin{eqnarray}
L &\equiv &\zeta ^{\mu }u_{\mu }=-\frac{2Ma\rho \sin ^{2}\theta }{\Sigma }%
\overset{\cdot }{t}  \notag \\
&&+\frac{\left( \rho ^{2}+a^{2}\right) ^{2}-\Delta a^{2}\sin ^{2}\theta }{%
\Sigma }\sin ^{2}\theta \overset{\cdot }{\phi },  \label{ker-l}
\end{eqnarray}%
where again $\overset{\cdot }{t}=\frac{dt}{ds}\equiv u^{t}$ and $\overset{%
\cdot }{\phi }=\frac{d\phi }{ds}\equiv u^{\phi }$. In addition, the Kerr
metric admits a symmetric second-rank Killing tensor $K_{\mu \nu }$ that
satisfies the Killing equation $K_{\mu \nu ;\alpha }+K_{\mu \alpha ;\nu }=0$
and generates a quadratic integral of motion expressed by the $4-$scalar $%
K\equiv K_{\mu \nu }u^{\mu }u^{\nu }$. This is expressed as%
\begin{equation}
K\equiv \mathcal{Q}+\left( L-aE\right) ^{2},  \label{K}
\end{equation}%
which can be shown to be always non-negative and where $\mathcal{Q}$ denotes
the Carter constant%
\begin{equation}
\mathcal{Q=}p_{\theta }^{2}+\cos ^{2}\theta \left[ a^{2}\left(
1-E^{2}\right) +\left( \frac{L}{\sin \theta }\right) ^{2}\right] ,
\end{equation}%
and here $p_{\theta }=\Sigma \overset{\cdot }{\theta }$, with $\overset{%
\cdot }{\theta }=\frac{d\theta }{ds}\equiv u^{\theta }$. The invariant $K$
can be represented as a polynomial of the particle velocity components as%
\begin{equation}
K=\Sigma ^{2}u_{\theta }^{2}+A_{1}u_{t}^{2}+A_{2}u_{\phi
}^{2}+A_{3}u_{t}u_{\phi }+A_{4},  \label{k-poli}
\end{equation}%
where $A_{1}$, $A_{2}$, $A_{3}$ and $A_{4}$ are configuration-space
functions given respectively by%
\begin{eqnarray}
A_{1} &\equiv &a^{2}\sin ^{2}\theta , \\
A_{2} &\equiv &\left( \rho ^{2}+a^{2}\right) ^{2}\sin ^{2}\theta , \\
A_{3} &\equiv &-2a\rho \sin ^{2}\theta , \\
A_{4} &\equiv &a^{2}\cos ^{2}\theta ,
\end{eqnarray}%
and where, without possibility of misunderstanding, here the subscripts $1-4$%
\ are just labels not related to tensorial indexes. This type of dependence
on the particle $4-$velocity carried by $K$ is of crucial importance for the
occurrence of non-ideal fluid configurations in Kerr space-time through the
generation of temperature anisotropy.

The realization of relativistic kinetic equilibria for collisionless $N-$%
body systems of neutral matter in Kerr space-time follows from the
identification of the invariants with the set $I\left( \mathbf{x}\left(
s\right) \right) =(E,L,K)$ defined respectively by Eqs.(\ref{ker-e}), (\ref%
{ker-l}) and (\ref{K}). The equilibrium KDF is represented in the form $%
f=f_{\ast }$, where now%
\begin{equation}
f_{\ast }=f_{\ast }\left( E,L,K,\Lambda _{\ast }\right)  \label{f-star-kerr}
\end{equation}%
is still a smooth strictly-positive function of the particle invariants
only, which is summable in velocity-space. The notation is analogous to that
introduced in Eq.(\ref{f-star}). In particular, the structure functions $%
\Lambda _{\ast }$ carry implicit functional dependences on the invariants $%
I\left( \mathbf{x}\left( s\right) \right) $, so that their general
representation is of the type $\Lambda _{\ast }=\Lambda _{\ast }\left(
E,L,K\right) $, although for this analysis it is convenient to restrict to
the particular case in which $\Lambda _{\ast }$ are identically constant,
namely $\Lambda _{\ast }=const$. An explicit representation of the
equilibrium KDF $f_{\ast }$ given by Eq.(\ref{f-star-kerr}) and holding in
Kerr metric and carrying linear powers on the invariants can be prescribed
in terms of a Gaussian-like distribution as%
\begin{equation}
f_{\ast }=\beta _{\ast }e^{-E\gamma _{\ast }-L\omega _{\ast }-K\alpha _{\ast
}},  \label{gauss-kerr}
\end{equation}%
generally different from a Maxwellian solution. In this case the structure
functions $\Lambda _{\ast }=\left( \beta _{\ast },\gamma _{\ast },\omega
_{\ast },\alpha _{\ast }\right) $ are related to the system physical
observables according to following prescription: $\beta _{\ast }$ is
associated with the system number density measured in the fluid comoving
frame, $\gamma _{\ast }$ determines the system isotropic temperature, $%
\omega _{\ast }$ defines the fluid angular frequency along the direction $%
\phi $, when measured by an inertial observer, and, finally, $\alpha _{\ast
} $ is peculiar of the Kerr solution, being related to the system
temperature (and pressure) anisotropy. We notice that the role of the
invariant $K$ in the Gaussian-like KDF $f_{\ast }$ (\ref{gauss-kerr}) is
analogous to that of particle magnetic moment $m^{\prime }$ in relativistic
plasmas and responsible for the occurrence of a temperature anisotropy
effect in those systems \cite{pop2}. In fact, the function $\alpha _{\ast }$
determines a phase-space anisotropy contribution carried by the constant $K$%
, so that it generates a deviation from the isotropic case.

Under the assumption $\Lambda _{\ast }=const.$, the non-isotropic character
of $T^{\mu \nu }\left( r\right) $ follows now as a unique consequence of the
dependence of $f_{\ast }$ on the invariant $K$ characteristic of the
relativistic equilibrium solution in the Kerr metric. In fact, thanks to Eq.(%
\ref{k-poli}), the integrand in the previous equation becomes a polynomial
function carrying a non-isotropic dependence on the particle velocity
components $\left( u_{\phi },u_{\rho },u_{\theta }\right) $, while the
configuration-space functions $A_{1}-A_{4}$ are all different from each
other. In addition, the metric tensor of the Kerr space-time is
non-diagonal. As a result of the combination of these features, two kind of
deviations from the ideal fluid case (described by the isotropic Maxwellian
KDF) arise: a) all the diagonal terms of the tensor $T^{\mu \nu }\left(
r\right) $ are generally different from each other; b) $T^{\mu \nu }\left(
r\right) $ carries also non-vanishing off-diagonal terms generated by the
symmetric entries $T^{t\phi }\left( r\right) =T^{\phi t}\left( r\right) $
which arise due to the presence of the term $A_{3}u_{t}u_{\phi }$ in Eq.(\ref%
{k-poli}). The non-isotropic stress-energy tensor $T^{\mu \nu }\left(
r\right) $ associated with the non-ideal fluid then generates corresponding
non-isotropic directional temperatures, carrying a non-trivial dependence on
the space-time metric tensor components determined by explicit solution of
the integrals (\ref{tmunu-bis}).

The resulting configuration is therefore substantially different from the
ideal fluid described by the isotropic Maxwellian KDF in Schwarzschild
space-time, for which the Tolman-Ehrenfest effect applies. This proves that
the same relation ceases to hold when either the space-time does not satisfy
the diagonal representation condition or the ideal fluid hypothesis is
violated. In the case of collisionless matter at kinetic equilibrium in Kerr
space-time these two effects show up simultaneously, mining the validity of
the Tolman-Ehrenfest law. A final discussion is also in order concerning the
connection between the solutions in Kerr and Schwarzschild space-times. In
fact, the constant of motion $K$\ defined by Eq.(\ref{K}) for the Kerr
metric naturally survives also for the Schwarzschild geometry. Therefore,
the equilibrium KDF of the type (\ref{f-star-kerr}) can be a general
representative of non-isotropic kinetic solution for both space-times. In
fact, the Schwarzschild configuration can be recovered as limiting case by
letting the parameter $a\equiv S/M\rightarrow 0$, preserving the
non-isotropic character of the temperature. As a result, violation of the
Tolman-Ehrenfest law due to the Carter constant is in principle achievable
also in the Schwarzschild space-time, for a non-Maxwellian fluid described
by the non-isotropic KDF (\ref{gauss-kerr}).

In conclusion, the physical cases discussed here are potentially interesting
in astrophysics, as they include realistic features beyond the ideal-fluid
solution. It is therefore conjectured that in real physical systems arising
around compact objects it is reasonable to expect a violation of the
Tolman-Ehrenfest relation (\ref{2}), to be replaced by more involved
implicit predictions of non-isotropic temperature profiles given by the
directional integrals (\ref{tmunu-bis}) for corresponding kinetic equilibria.

\textbf{AUTHOR'S CONTRIBUTIONS}

All authors contributed equally to this work.

The authors have no conflicts to disclose.

\textbf{ACKNOWLEDGMENTS}

CC, JK and ZS acknowledge the support of the Research Centre for Theoretical
Physics and Astrophysics, Institute of Physics, Silesian University in
Opava, Czech Republic.

\textbf{DATA AVAILABILITY}

The data that supports the findings of this study are available within the
article.

\end{document}